\documentclass[%
 reprint,
 amsmath,amssymb,
 aps,
]{revtex4-2}

\usepackage{graphicx}
\usepackage{dcolumn}
\usepackage{bm}
\usepackage{amsmath}
\usepackage{comment}
\usepackage[dvipsnames]{xcolor}
\usepackage{subfigure}
\usepackage[colorlinks]{hyperref} 

\begin{document}

\preprint{APS/123-QED}

\title{Coherent control phase lag across doubly-excited atomic strontium resonances in an $\omega-2\omega$ interference scheme}

\author{Yimeng Wang}
\altaffiliation[Present address: ]{Fachbereich Physik, Freie Universität Berlin, Arnimallee 14, D-14195 Berlin, Germany.}

\author{Chris H. Greene}%

\affiliation{Department of Physics and Astronomy, Purdue University, West Lafayette, Indiana 47907, USA and
Purdue Quantum Science and Engineering Institute,
Purdue University, West Lafayette, Indiana 47907, USA}

\date{\today}

\begin{abstract}
Accurate calculations of phase lag associated with coherent control where an excited system decays into more than one product channel have recently been reported for atomic barium in PhysRevA.105.013113 (2022). The present study extends the calculations to make predictions of that observable for another alkaline earth atom, strontium, with a discussion of its spectrum and phase lag for energies between the $Sr^{+}$ $4d_{3/2}$ and $4d_{5/2}$ thresholds. We will explore the physics influenced by the electron correlations of strontium and the long-range Coulomb potential. 
The behavior of the phase lag cannot be simply addressed by a time delay analysis, although the latter is often used to address the prominent channel of resonance decay of doubly excited states. 
\end{abstract}

\maketitle
\section{Introduction}

Coherent control in atoms and molecules, especially relating to the control of phase lag, has been studied for decades in both experiments \cite{PhysRevA.76.053401,PhysRevLett.98.053001,Zhu77,PhysRevLett.92.113002,PhysRevLett.74.4799,PhysRevLett.96.173001,PhysRevLett.79.4108} and theory \cite{PhysRevLett.86.5454,PhysRevLett.79.4108,Nakajima_1997,PhysRevLett.82.2266,PhysRevLett.82.65,PhysRevA.50.595,Seideman:1999,PhysRevLett.70.1081,PhysRevLett.69.2353,Anderson:1992,PhysRevA.100.063417,PhysRevX.10.031070,PhysRevLett.82.4791,PhysRevA.100.043409,PhysRevA.103.053118,PhysRevA.105.013113}.
Our recent article \cite{PhysRevA.105.013113} presented calculations of the phase lag between the $6s_{1/2}$ and $5d_{3/2}$ continua in atomic barium using a streamlined variational $R$-matrix method and multichannel quantum defect theory (MQDT), and obtained generally good agreement with experiments \cite{PhysRevA.76.053401,PhysRevLett.98.053001}, improving significantly on the results of a simplified phenomenological model that has been widely discussed in previous theoretical literature \cite{PhysRevLett.86.5454,PhysRevLett.79.4108,Nakajima_1997,PhysRevLett.82.2266,PhysRevLett.82.65,PhysRevA.50.595,Seideman:1999,PhysRevLett.70.1081}. Our results in Ref. \cite{PhysRevA.105.013113} demonstrated the capability of accurate, nearly {\it ab initio} calculations of the complex heavier alkaline earth atoms to describe such phenomena. The present study extends those calculations to another alkaline earth atom, strontium, which turns out to differ significantly from barium, and we develop predictions that can be tested in future experimental studies of the phase-controlled interference.  Strontium is another system with two valence electrons whose spectrum has been extensively calculated and measured \cite{Aymar_1987,Kompitsas_1990,Kompitsas_1991,Goutis_1992,PhysRevA.44.1773,PhysRevA.46.1248,PhysRevA.49.1029,Vidolova-Angelova1994,Cohen2001}. 
Despite the differences in the core electron structure between barium and strontium, the logic and discussion in the present study follow the development in Ref. \cite{PhysRevA.105.013113}.

\begin{figure}[htbp]
  \includegraphics[scale=0.38]{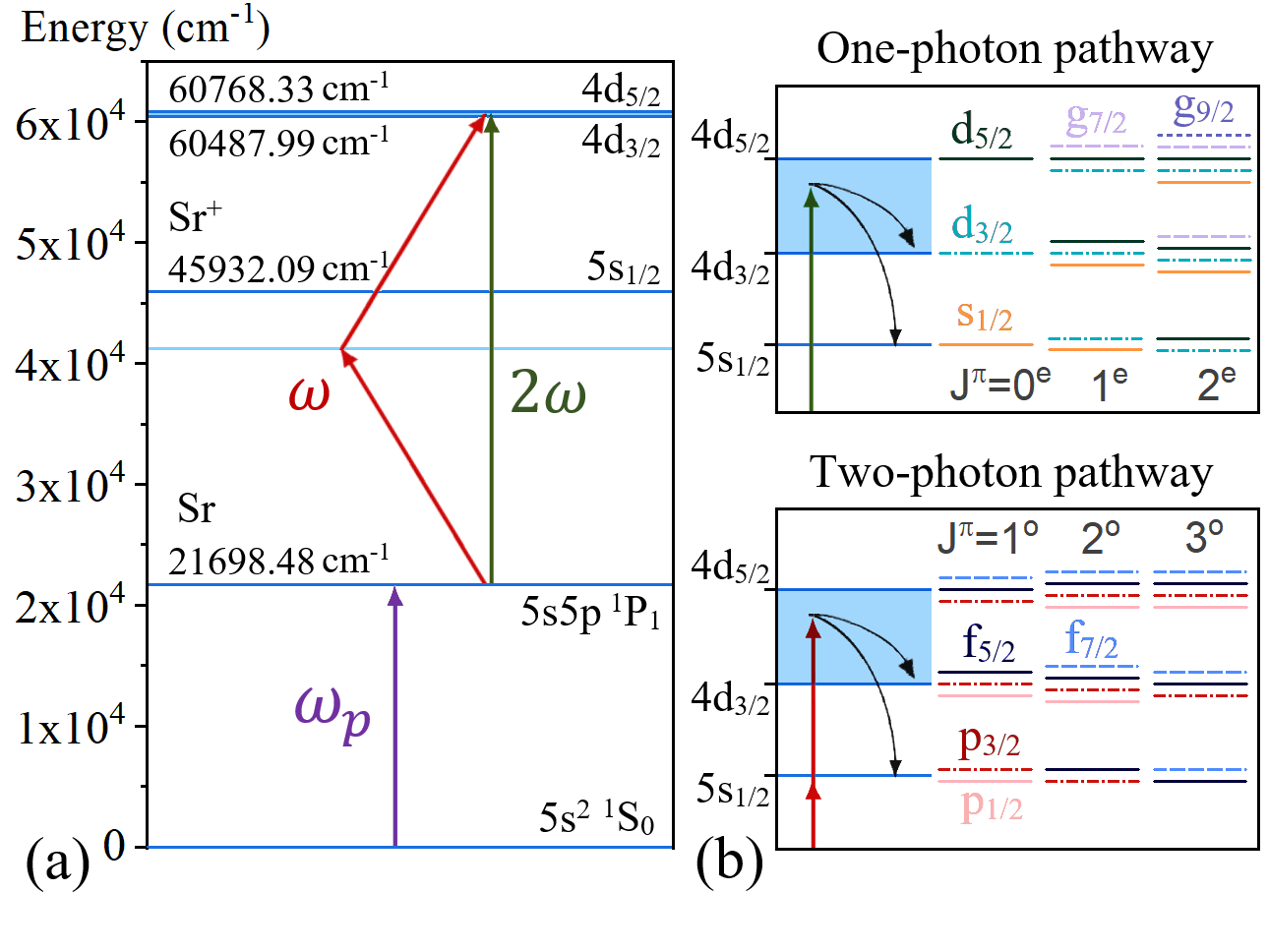}
    \caption{ (a) Energy level diagram for the strontium $\omega-2\omega$ interference scheme. Start with a ground state strontium $i =5s^2$, and first use a linearly-polarized pump laser ($\omega_p$) to excite the atom to the $e_1=5s5p$ $(^1P_1)$ state. Next, when the atom is ionized through simultaneous absorption of two photons of frequency $\omega$ and one photon of frequency $2\omega$, the final state reaches the shaded energy region, which can decay into the continua associated with either the $5s_{1/2}$ or the  $4d_{3/2}$ ionic state. 
    (b) Channels relevant to autoionizing decay via either the one- or two-photon pathway for each total angular momentum and parity $J^{\pi}$. The possible orbital angular momenta $l_j$ of the photoelectron in each continuum are labeled by the lines, distinguished by different colors and line types. Analogous electronic structure is displayed in Table.I of  Ref. \cite{PhysRevA.105.013113} for Ba. 
    }
  \label{1}
\end{figure}

The ionization scheme to be applied can be found in Fig. \ref{1} (a). It is based on the $\omega-2\omega$ coherent phase control scenario, which has been widely used in many different regimes, such as physical chemistry \cite{Zhu77,PhysRevLett.92.113002,PhysRevLett.74.4799,PhysRevLett.96.173001}, condensed matter \cite{PhysRevLett.74.3596,PhysRevLett.92.147403}, nuclear physics \cite{nucl}, quantum computation \cite{PhysRevLett.111.233002} and ultrafast physics \cite{PhysRevLett.117.217601,PhysRevLett.107.153902}. 
The idea behind coherent phase control is to introduce a tunable optical phase difference $\Delta\Phi$ between the two laser fields, permitting a manipulation of the interference between them and thereby control over the observable outcomes, manifested as sinusoidal modulations as a function of $\Delta\Phi$, when the different optical routes lead coherently to the same final state.
The $\omega-2\omega$ interference scheme only controls observables that have odd parity, e.g. odd order Legendre polynomials, as a result of the photoabsorption selection rule that constrains the  one- and two- photon absorptions to reach states with opposite parities $\pi=\{e,o\}$ \cite{PhysRevA.105.013113}. 
This paper focuses on the photoelectron angular distribution (PAD) under the $\omega-2\omega$ coherent phase control scheme. Especially, we consider the situation where the final autoionizing states can decay into more than one continuum, leaving the residual ion in states of different energy. The phase lag between two continua $i,i^{\prime}$ is defined by $\Delta\delta=\delta^{(i)}-\delta^{(i^{\prime})}$. It has been found in experiments that the phase lag behaves sensitively in the vicinity of autoionizing resonances \cite{Seideman:1998,PhysRevLett.79.4108,PhysRevLett.82.65}, which serves as a major topic in the discussions below. 


This article is organized as follows: Sec. \ref{s-2} introduces the underlying theory and formulas (in parallel with Sec. II of Ref. \cite{PhysRevA.105.013113}). Sec. \ref{s-3}  presents the results and discussion: Sec. \ref{s-3.1} demonstrates the separate one- and two-photon ionization cross sections ignoring the interference between the pathways; Sec. \ref{s-3.2} shows the phase lag, as the major observable reflecting the coherent phase control between one- and two-photon pathways; Sec. \ref{s-3.3} provides a time-delay analysis of the decay channels of strontium doubly excited states. Finally, Sec. \ref{s-4} summarizes our main conclusions.

\begin{figure*}[htbp]
  \includegraphics[scale=0.35]{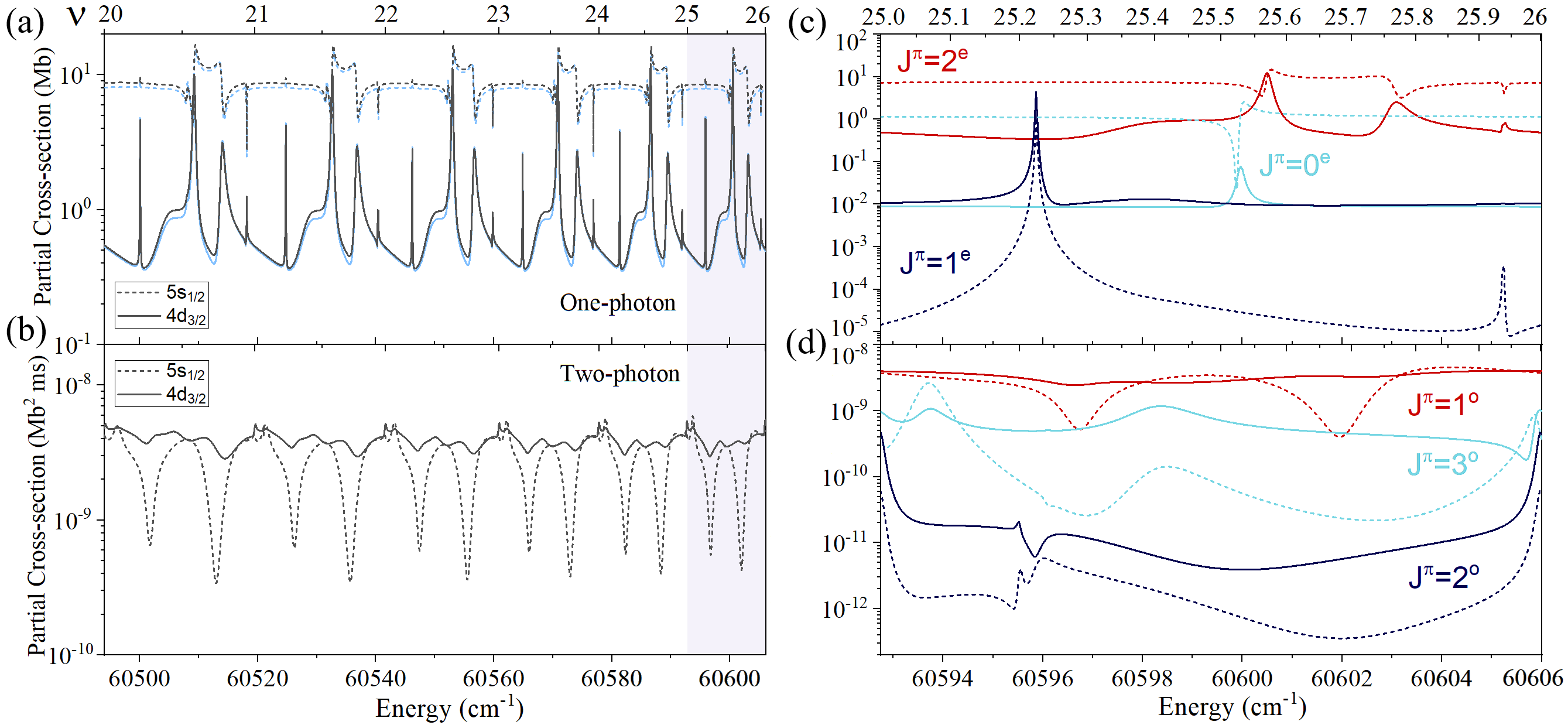}
  \caption{Partial cross sections for one- (upper-) and two-photon (lower-panels) ionization. The dashed (solid) curves give the components for $5s_{1/2}$($4d_{3/2}$) continuum for (a-d). Results in (a) are calculated in both length and velocity gauge \cite{LAMBROPOULOS1998203,Burke_1990}, shown in the dark and light colors respectively. Results in (b)-(d) are calculated only in the length gauge. 
  Subfigure (a,b) displays a wider energy range of cross sections just above $4d_{3/2}$ threshold, while (c,d) shows each $J^{\pi}$ partial cross sections for only one principal quantum number cycle (the shadowed region in left panel). 
  Resonance identifications and their comparisons with experiment are presented in Table. \ref{tab:table4}. The energy scales for all the figures in this paper are given in $cm^{-1}$ in the bottom sub-figures, with the effective quantum number $\nu=[2(E_{4d_{5/2}}-E)]^{-\frac{1}{2}}$ labeled in the top sub-figures. }
  \label{3}
\end{figure*}

\section{Method}\label{s-2}

The ionization steps shown in Fig. \ref{1} (a) work as follows: a ground state strontium atom is pumped firstly onto $e_1=5s5p$ $(^1P_1)$ state, then reaches final states in the shadowed region by a concurrent $\omega-2\omega$ ionization.
Now viewing the $e_1=5s5p$ $(^1P_1)$ $M_J=0$ state as the initial state of the system, the Hamiltonian for the $\omega-2\omega$ ionization becomes,

\begin{equation}
\label{eq1-1}
\begin{split}
    &H(\vec{r}_1,\vec{r}_2,t)=\sum_{i=1,2}\left( -\frac{\nabla_{\vec{r}_i}^2}{2} -\frac{\tilde{Z}(r_i)}{r_i} + \vec{\mathcal{E}}(t)\cdot \vec{r}_i\right)+\frac{1}{|\vec{r}_1-\vec{r}_2|} \\
    &\vec{\mathcal{E}}(t)=\frac{\hat{z}}{2}\left(\mathcal{E}_{\omega} e^{-i\omega t}+\mathcal{E}_{2\omega} e^{-i(2\omega t-\Delta\Phi) }+ c.c.\right)
\end{split}
\end{equation}

The strontium energy eigenstates in this energy range are described adequately by just the two outermost active electrons ($\vec{r}_1,\vec{r}_2$). $\tilde{Z}(r_i)$ is a semiempirical parameter that describes the model potential of $Sr^{2+}$ \cite{Review:1996}. This model potential is determined by the core electrons, and it includes an $l_i$-dependent screening term and a polarization term with a short-range cutoff. Compared to atomic barium, strontium has fewer core electrons and therefore a smaller screening and core polarization effect, which leads to their significantly different energy structures. No dielectronic polarization term is included in the Hamiltonian, as its main effect is to make the Hamiltonian slightly more repulsive for configurations where both electrons come close to the ionic core at the same time, such as in the ground state \cite{Review:1996}. When studying photoionization spectra, it has only a minor effect, as demonstrated extensively by the literature on many alkaline earth atomic spectra \cite{PhysRevA.36.4272,PhysRevA.36.2706,PhysRevA.44.1773,PhysRevA.47.229}. 
Note that $\vec{\mathcal{E}}(t)$ is the total external electric field at the position of the atom, and a controllable phase difference $\Delta\Phi$ has been introduced between its fundamental and second-harmonic components. All the laser beams depicted in Fig. \ref{1} (a) are linearly polarized along $\hat{z}$, preparing the system with azimuthal symmetry. The photon-atom interaction is given in the length gauge. 
The quantum numbers for the inner and outer electron are denoted by $(s_c, L_{c}){J_c}$ and $(s,l)j$, respectively.  $\vec{J_{cs}}=\vec{J_{c}}+\vec{s}$, $\vec{J}=\vec{J_{cs}}+\vec{l}$. Every possible asymptotic quantum state after ionization has been listed in Fig. \ref{1} (b) . The photoelectrons can leave the ion from either $n_c L_{c_{J_c}}=5s_{1/2}$ or $n_c L_{c_{J_c}}=4d_{3/2}$ threshold with different kinetic energies, their angular momenta $J^{\pi}(l_j)$ obey the selection rules and therefore be different for one- and two-photon pathways.  

By treating the atom-photon interactions perturbatively, the cross section for the $q$-photon ionization process ($q=1,2$) is:
\begin{equation}
\label{eq2-3}
\begin{split}
   \sigma_{e_1\rightarrow f}^{(q)}=2\pi(2\pi\alpha\omega)^{q}\left|T_{f,e_1}^{(q)}\right|^2
\end{split}
\end{equation}
$T_{f,e_1}^{(q)}$ is the $q$-photon transition amplitude from initial state $e_1$ to final state $f$ ($\vec{r}=\vec{r}_1+\vec{r}_2$, $e_2$ denotes the intermediate states):
\begin{equation*}
\begin{split}
T_{f,e_1}^{(q)}&=
\begin{cases}
 \langle [J_{c},(s,l)j]J_f M_f|r_{z}|J_{e_1} M_{e_1}\rangle& q=1 \\ 
 &\\
 \sum_{e_2}\frac{\langle [J_{c},(s,l)j]J_f M_f|r_{z}|e_2\rangle 
 \langle e_2|r_{z}|J_{e_1} M_{e_1}\rangle}{E_{e_2}-E_{e_1}-\omega}& q=2   \\
\end{cases}\\
\end{split}
\end{equation*}

The spin precession of the initial state $e_1$ will produce the ``hyperfine depolarization" effect. 
Although $93\%$ natural strontium have nuclear spin $I=0$, there are remaining $7\%$ $I=\frac{9}{2}$ isotopes \cite{Elements} that cause hyperfine splittings and make $F, M_F$ ($\vec{F}=\vec{J}+\vec{I}$) the conserved quantum number. A detailed discussion and formalism for treating this phenomenon can be found in Ref. \cite{PhysRevA.105.013113}. For our system where all the photons are linearly-polarized, it introduces parity-unfavored [$(-1)^{l_1+l_2}\neq (-1)^L$] partial waves, and changes the cross section from Eq. \ref{eq2-3} into, 
\begin{equation}
\label{eq2-2}
\begin{split}
   &\sigma_{e_1\rightarrow f}^{(q,hfp)}%
   =\sum_{M_{J_{e_1}}=0,\pm1}\sigma_{e_1\rightarrow f}^{(q)}\left[\frac{1}{3}+\left(\frac{2}{3}-|M_{J_{e_1}}|\right)g^{(2)}_{ave} \right]\\
\end{split}
\end{equation}

The definition of the constant $g^{(2)}_{ave}$ can be found in Ref. \cite{PhysRevA.47.229}. $g^{(2)}_{ave}=1$ indicates no hyperfine effects. In this paper we take $g^{(2)}_{ave}=0.9446$, which is obtained by averaging over the isotopes and taking the long time limit. It only causes a tiny correction to the spectrum in the energy range considered here.

The PADs under concurrent $\omega-2\omega$ ionization are quantified by the differential ionization rate $\frac{dW}{d\Omega}$ as below \cite{Lindsay:1992}. Since the PADs are detected separately for photoelectrons with different kinetic energies, $\frac{dW}{d\Omega}$ is calculated separately for the two specific groups of photoelectrons that leave the ion in either the $(5s_{1/2})$ or $(4d_{3/2})$ state. Their distinguishing index is currently omitted for brevity, until later when we compare the PADs in those two different continua. 

\begin{equation*}
\begin{split}
\frac{dW}{d\Omega}=&\mathcal{N}^2\sum_{\lambda_i}
\bigg| \mathcal{E}_{\omega}^2\sum_{l=odd} Y_{l m}(\theta,\phi) f_{\gamma,2}e^{i\delta_{\gamma,2}} \\
&+ \mathcal{E}_{2\omega} e^{i \Delta\Phi}\sum_{l=even} Y_{l m}(\theta,\phi) f_{\gamma,1}e^{i\delta_{\gamma,1}}
\bigg|^2\\
\end{split}
\end{equation*}
\begin{equation}
\label{eq-1}
\begin{split}
f_{\gamma,q}e^{i\delta_{\gamma,q}}=&\sum_{\lambda_c}
\langle J_{cs}M_{J_{cs}},l m|J_f,M_{J_f}\rangle  \\
&\times\langle [(s,J_{c})J_{cs},l]J_f|[J_{c},(s,l)j]J_f\rangle T_{f,i}^{(q)}.
 \end{split}
\end{equation}

$\mathcal{N}^2$ is a normalization constant. $T_{f,i}^{(q)}$ is the transition amplitude from ground state $i$ to final state $f$, whose relation to $T_{f,e_1}^{(q)}$ can be found in the Appendix of Ref. \cite{PhysRevA.105.013113}. The incoherent sum index ${\lambda_i}$ includes $J_{cs}$, $I$, and all the angular momentum projections $M$. The coherent sum index represents ${\lambda_c}=\{J_f,j\}$, and $\gamma=\{\lambda_i,l\}$. 


To study the interference between one- and two-photon ionization pathways, which influences the asymmetry of the PAD, consider the directional asymmetry parameter $\alpha_{asym}$, defined as the ratio between the $-z$ directed photoelectron current and the total photocurrent \cite{PhysRevLett.98.053001,PhysRevA.76.053401,PhysRevA.103.053118,PhysRevA.105.013113}:
\begin{equation}
\label{eq-2}
    \alpha_{asym}= \frac{2\pi}{W_{\rm tot}} \int_{\frac{\pi}{2}}^{\pi} \frac{dW(\theta)}{d\Omega}\sin \theta d\theta
    \equiv\frac{1}{2}+A \cos{(\Delta\Phi-\delta_0)} 
\end{equation}
$\alpha_{asym}$ can be parameterized by an amplitude $A$ and phase $\delta_0$. $0\leq A\leq \frac{1}{2}$ and $0\leq\delta_0\leq2\pi$. The electric field strengths $\mathcal{E}_{\omega,2\omega}$ affect $A$ only. The phase $\delta_0$ can be expressed in terms of the complex coefficients $fe^{i\delta}$ and angular integrals as: 
\begin{equation}
\label{eq-3}
\begin{split}
    \delta_0&=-arg\left[\sum_{\lambda_i,l_o,l_e,k_o}\rho_{k_o}\Theta(l_o,l_e,k_o,m) f_{\gamma_o}f_{\gamma_e}e^{i(\delta_{\gamma_e}-\delta_{\gamma_o})}\right]
\end{split}    
\end{equation}
where 
\begin{small}
\begin{equation*}
\begin{split}
&\rho_k=\int_{-1}^{0} P_k(x) dx \\
&\Theta(l,l^{\prime},k,m)=(2k+1)\int Y_{l,m}(\theta,\phi) P_{k}(\cos{\theta}) Y_{l^{\prime},m}^{*}(\theta,\phi) d\Omega
\end{split}    
\end{equation*}
\end{small}

$P_k(x)$ is the $k$-th order Legendre polynomial. Note that the subscript $e$($o$) for $l$ and $k$ denotes even(odd) numbers, and $\gamma_{e(o)}=\{\lambda_i,l_{e(o)}\}$. Especially, we will be interested in the difference between the $\delta_0$ from different continua, which is the so called ``phase lag" $\Delta\delta= \delta_0^{(5s_{1/2})}-\delta_0^{(4d_{3/2})}$. The calculated cross sections, PADs and the phase lag based on Eqs. \ref{eq2-2}, \ref{eq-1} and \ref{eq-3} are presented in the next section.

\section{Results and discussion}\label{s-3} 

The calculations to be shown in this section are based on the multichannel quantum defect theory (MQDT) \cite{Greene:1979,Seaton:1983,GRF1982,Greene:1985} and the variational streamlined $R$-matrix method \cite{Review:1996}. It is assumed that the electron correlations are restricted to the region where both electrons are inside a reaction zone, where the wave function can be described by anti-symmetric products of two-electron basis functions $\mathcal{Y}(n_c L_{c_{J_c}};nl_j)$. 
According to the energy levels and electron structure of strontium, the reaction zone is set to be within 70 $a.u.$ from the nucleus, to include the possible influences of $5s_{1/2}$ Rydberg series to the intermediate states, and $4p_{1/2,3/2}$ Rydberg series to the final states. $L_c,l = 0-4$, and the radial basis functions have at most 60 nodes.
The validity of our calculations has been confirmed by comparing with previous experimental observations \cite{Kompitsas_1990,Kompitsas_1991,Goutis_1992} in Table. \ref{tab:table4}. In the Appendix the quantum defect matrix is tabulated for each symmetry, which can potentially be useful for future numerical studies. 

\subsection{Photoelectron spectrum and the angular distributions with no interference }\label{s-3.1}

To begin with, we consider the case where no interference exists between the $\omega-2\omega$ pathways. The cross sections are given in Fig. \ref{3}. In subfigures (a), (b), the final state energy is roughly from 60494 to 60606 $cm^{-1}$ relative to the ground state, which lies just above the $4d_{3/2}$ threshold. The effective principal quantum number of the fragmentation electron relative to the $4d_{5/2}$ threshold is $\nu=1/[2(E_{4d_{5/2}}-E)]^{\frac{1}{2}}=20-26$. A comparison between results using length and velocity gauges \cite{LAMBROPOULOS1998203,Burke_1990} is given in subfigure \ref{3} (a) as a check on the accuracy of the calculations. 
The photoelectron can be separated into a fast ($5s_{1/2}$) and a slow ($4d_{3/2}$) continuum depending on their kinetic energies, and their cross sections are shown in dashed and solid lines, separately. For the one-photon cross sections, it is the $5s_{1/2}$ continuum that dominates; while for the two-photon cross sections, it is the $4d_{3/2}$ continuum that dominates. 
To resolve the details of the resonance structures, we show in subfigures (c), (d) the cross sections of different $J^{\pi}$-partial waves within one cycle of $4d_{5/2}$ Rydberg series (the shadowed region). Among the one-photon pathways, the $J^{\pi}=2^e$ partial wave is dominant for both the $5s_{1/2}$ and $4d_{3/2}$ continua; while among the two-photon pathways, it is the $J^{\pi}=1^o$ partial wave that dominates for both continua. The parity-unfavored partial waves $J^{\pi}=1^e$ and $2^o$, present only because of hyperfine depolarization, have the smallest contributions except at their resonance peaks ($\nu=25.23\,(1^e)$ and $\nu=25.00\,(2^o)$). 
The positions of resonances and their identifications can be found in Table. \ref{tab:table4}. Some of the resonances with $J^{\pi}=0^e, 1^e, 2^e$ and $3^o$ have been observed by previous experiments \cite{Kompitsas_1990,Kompitsas_1991,Goutis_1992}, and for the resonances that have been previously reported, our predicted energies are in most cases within the observed widths. For effective principle quantum numbers $\nu\approx 25$, the quantum defects should be close to being constants. However, it is not true for the spectrum in $J^{\pi}=1^o$ symmetry, which do not show a periodic behavior within the energy range of this study, as a possible consequence of the fact that there is a $J^{\pi}=0^e$ excited state lying close to the range of intermediate energies.

\begin{table*}
\caption{\label{tab:table4}The resonances from our calculation and earlier experiments, identified within one Rydberg cycle ($\nu=25-26$) and given in $cm^{-1}$. The classifications of the resonances are in the $J_f(n l_j)$ basis. The resonance energies shown in bold front can be compared with the tabulated experimental observations \cite{Kompitsas_1990,Kompitsas_1991,Goutis_1992}. }
\begin{ruledtabular}
\begin{tabular}{c|ccccc}
\multicolumn{6}{c}{Theoretical identification of resonances $4d_{5/2}nl_{j}$ }\\
 \hline
$J^{\pi}=0^e$ & \textbf{60600.0}  ($d_{5/2}$) &  &  &  \\
$J^{\pi}=1^e$ & \textbf{60595.9} ($d_{5/2}$)& 60598.4 ($d_{3/2}$)& 60605.2 ($g_{7/2}$)&   &   \\
$J^{\pi}=2^e$ & 60597.9 ($d_{3/2}$)& 60600.5 ($d_{5/2}$)& \textbf{60603.0} ($s_{1/2}$)& \textbf{60605.2} ($g_{7/2}$)& \textbf{60605.3} ($g_{9/2}$)\\
$J^{\pi}=1^o$ & 60596.7 ($p_{3/2}$)& 60599.4 ($f_{5/2}$)& 60602.4 ($f_{7/2}$)&  \\
$J^{\pi}=2^o$ & 60595.5 ($p_{3/2}$)& 60595.9 ($p_{1/2}$)& 60596.2 ($f_{7/2}$)& 60606.0 ($f_{5/2}$)&   \\
$J^{\pi}=3^o$ & \textbf{60593.7} ($f_{7/2}$)& 60596.1 ($p_{1/2}$)& \textbf{60598.3} ($p_{3/2}$)& \textbf{60605.9} ($f_{5/2}$)&  \\
 \hline  
\multicolumn{6}{c}{Experimental identification of resonances $4d_{5/2}nl_{j}$ (position E [width $\Gamma$])}\\
 \hline
$J^{\pi}=0^e$ \cite{Kompitsas_1991}& 60600.0 [0.3] &  &  &  \\
$J^{\pi}=1^e$ \cite{Goutis_1992}& 60596.1 [0.4]($27d_{5/2}$) &  &  &  \\
$J^{\pi}=2^e$ \cite{Goutis_1992}& 60603.1 [-]($29s_{1/2}$) & 60605.6 [-]($26g_{7/2}$) & 60605.6 [-]($26g_{9/2}$) &  &  \\
$J^{\pi}=3^o$ \cite{Kompitsas_1990}& 60593.9 [1.1] & 60599.4 [2.5] & 60606.0 [0.3] &  &  \\
\end{tabular}
\end{ruledtabular}
\end{table*}

\begin{figure}[htbp]
  \includegraphics[scale=0.32]{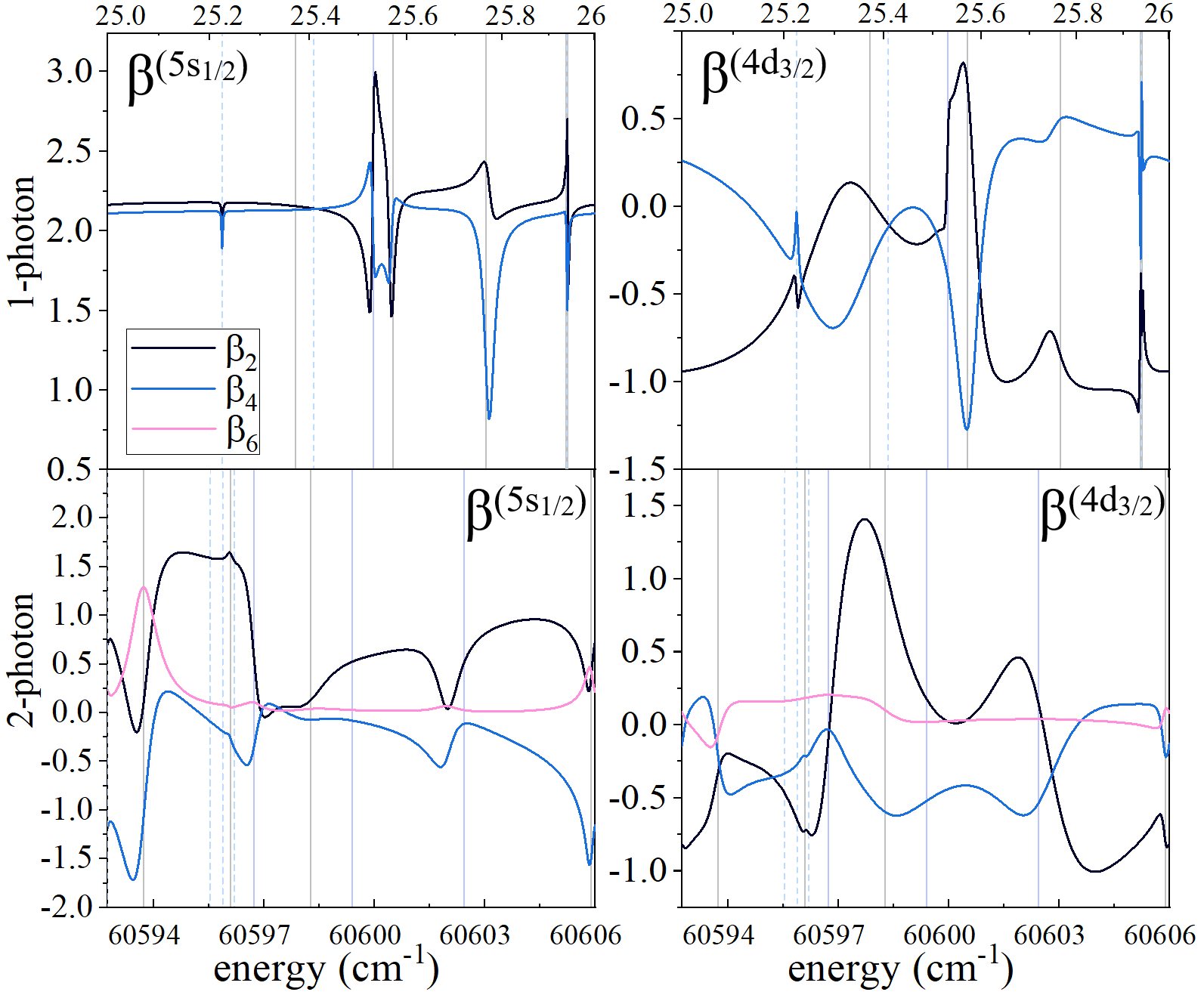}
    \caption{ The photoelectron angular distribution parameters $\beta_k^{(5s_{1/2},4d_{3/2})}$ for one- (upper-) and two-photon (lower-panel) ionization processes. $\beta_{2-6}$ are distinguished by different line colors. Without interference effect, the odd-order parameters $\beta_{k_o}=0$. 
    The left and right panels are $\beta$ parameters for two different continua $5s_{1/2}$ and $4d_{3/2}$. The vertical lines give the positions of the involved resonances [even- (odd-) parity resonances are for one (two)-photon processes]. }
  \label{4}
\end{figure}

Fig. \ref{4} demonstrates the PADs for one- and two-photon ionization separately, obtained by calculating the $\beta-$coefficients in Eq. \ref{eq2-4}.  Since all the lasers are linearly polarized along $\hat{z}$, the system has azimuthal symmetry, which allows one to arrange Eq. \ref{eq-1} into a sum of Legendre polynomials $P_k(\cos{\theta})$ as ($\theta$ is the polar angle between the direction of the ejected electron and the polarization axis $\hat{z}$):
\begin{equation}
\label{eq2-4}
\begin{split}
 \frac{dW}{d\Omega}&=\frac{W_{tot}}{4\pi}\sum_{k=0}^{6}\beta_k P_k(\cos{\theta})\\
\end{split}    
\end{equation}
For the one- or two-photon pathway, the maximum order of $k$ is 4 or 6, respectively. When there is no interference between the two pathways, the PADs are symmetrically distributed, and therefore all the odd order parameters $\beta_{k_o}$ are zero. The vertical lines give the positions of the involved resonances, where the PADs experience a dramatic change.

\subsection{Interference between one- and two-photon ionization pathways}\label{s-3.2}

\begin{figure}[htbp]
  \includegraphics[scale=0.35]{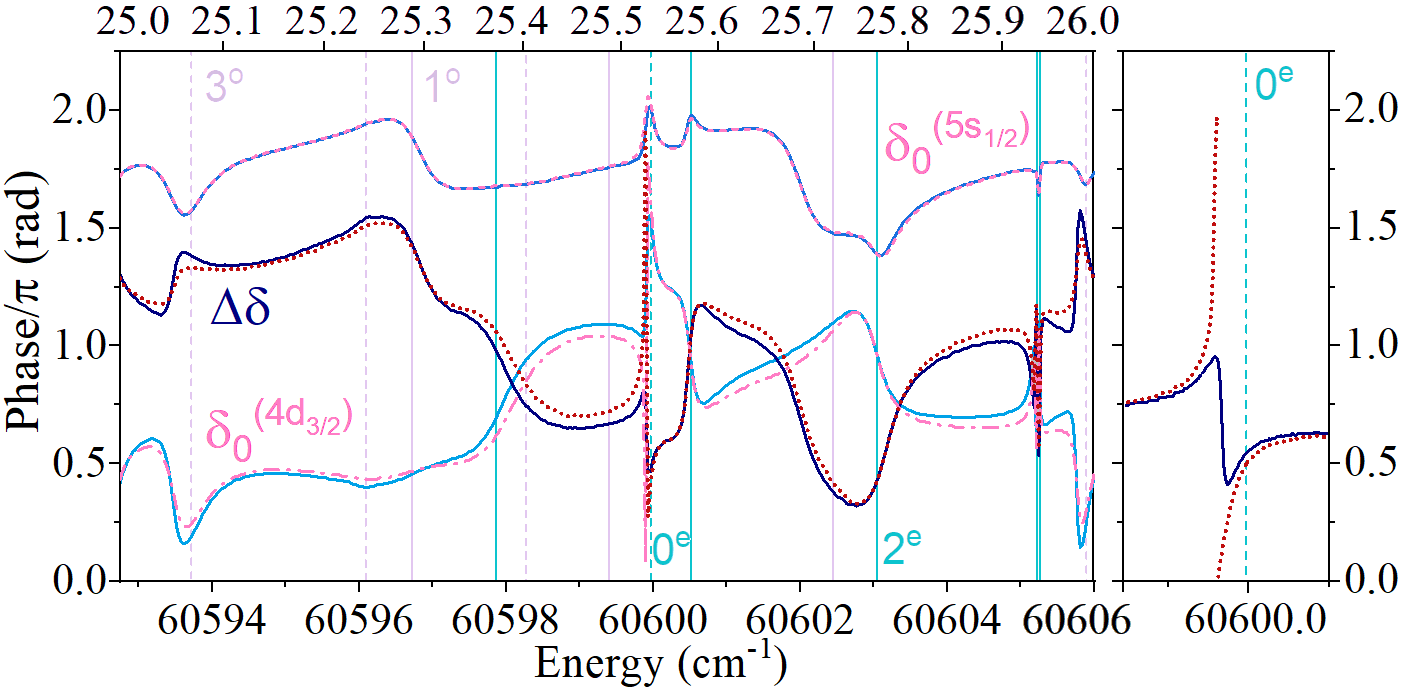}
    \caption{
    Left panel: Calculated phase lag $\Delta\delta=\delta_0^{(5s_{1/2})}-\delta_0^{(4d_{3/2})}$ and interference phase $\delta_0$ for $5s_{1/2}$ and $4d_{3/2}$  continuum. The solid and dashed/dotted curves give the results based on length and velocity gauge calculations, respectively. 
    The vertical lines give the positions of the one- and two-photon resonances, whose $J^{\pi}$ values are given by the label next to the lines. 
    Right panel: An enlarged picture of $\Delta\delta$ near the $J^{\pi}=0^e$ resonance at 60600 $cm^{-1}$. Influenced by the $J^{\pi}=0^e$ resonance, $\Delta\delta$ calculated from length gauge (solid) shifts only by a small value, while $\Delta\delta$ calculated from velocity gauge (dotted) changes by $2\pi$.
    }
  \label{2}
\end{figure}

Having discussed the one- and two-photon ionizations individually, we now turn to the interference effect between them. As introduced in some other studies \cite{PhysRevA.103.053118,PhysRevA.105.013113}, it can be quantified by $\alpha_{asym}=\frac{1}{2}+A \cos{(\Delta\Phi-\delta_0)}$.
Fig. \ref{2} demonstrates the $\delta_0$ for $5s_{1/2}$ and $4d_{3/2}$ continuum separately, and their difference  $\Delta\delta=\delta_0^{(5s_{1/2})}-\delta_0^{(4d_{3/2})}$. Since the hyperfine depolarization effect is negligible to $\delta_0$s, only resonances from parity-favored symmetries are marked in Fig. \ref{2}. 
Compared to our earlier calculations of phase lag in atomic barium (Fig. 2 of Ref. \cite{PhysRevA.105.013113}), there are many new properties in Fig. \ref{2} as a result of the different energy structure. 
For example, in the barium spectrum, many resonances are overlapping on each other, therefore the influence of each resonance to $\Delta\delta$ is unclear. The behavior of $\Delta\delta$ can hardly be connected to the lineshapes of the resonance peaks \cite{PhysRevA.76.053401}, and cannot be correctly predicted by a simplified model based on the interference between the direct-ionization and resonance-mediated ionization pathways \cite{PhysRevLett.86.5454,PhysRevLett.79.4108,Nakajima_1997,PhysRevLett.82.2266,PhysRevLett.82.65,PhysRevA.50.595,Seideman:1999,PhysRevLett.70.1081}. 
In contrast, the resonances of strontium in the range we considered are well-separated and more evenly distributed (whose positions are given by the vertical lines), so one can readily discern the influence of a single resonance on the phase lag.  
Another important feature reported in Ref. \cite{PhysRevA.105.013113} is that the phase lag appears to be quite robust over small perturbations in the system, whereby results from experiments and calculations could be expected to agree with each other accurately.  
However, Fig. \ref{2} turns out to be quite sensitive to small variations of the short-range-physics parameters. For example, if the one-photon ionization is calculated through velocity gauge \cite{LAMBROPOULOS1998203,Burke_1990} rather than the length gauge, the cross sections will only change slightly [as shown in Fig. \ref{3} (a)], but it modifies distinctively $\delta_0^{(4d_{3/2})}$ and $\Delta\delta$, especially near the $J^{\pi}=0^e$ resonance. As we can see in the enlarged picture of Fig. \ref{2}, where $\Delta\delta$ experiences only a small variation based on the calculations in the length gauge but fluctuates by $2\pi$ based on the calculations in the velocity gauge. The origin of this sensitivity remains unclear, and the true behavior of $\Delta\delta$ near the $J^{\pi}=0^e$ resonance will hopefully emerge from future experimental observations.

\subsection{Time delay analysis and the Coulomb phase shift}\label{s-3.3}

\begin{figure}[htbp]
  \includegraphics[scale=0.33]{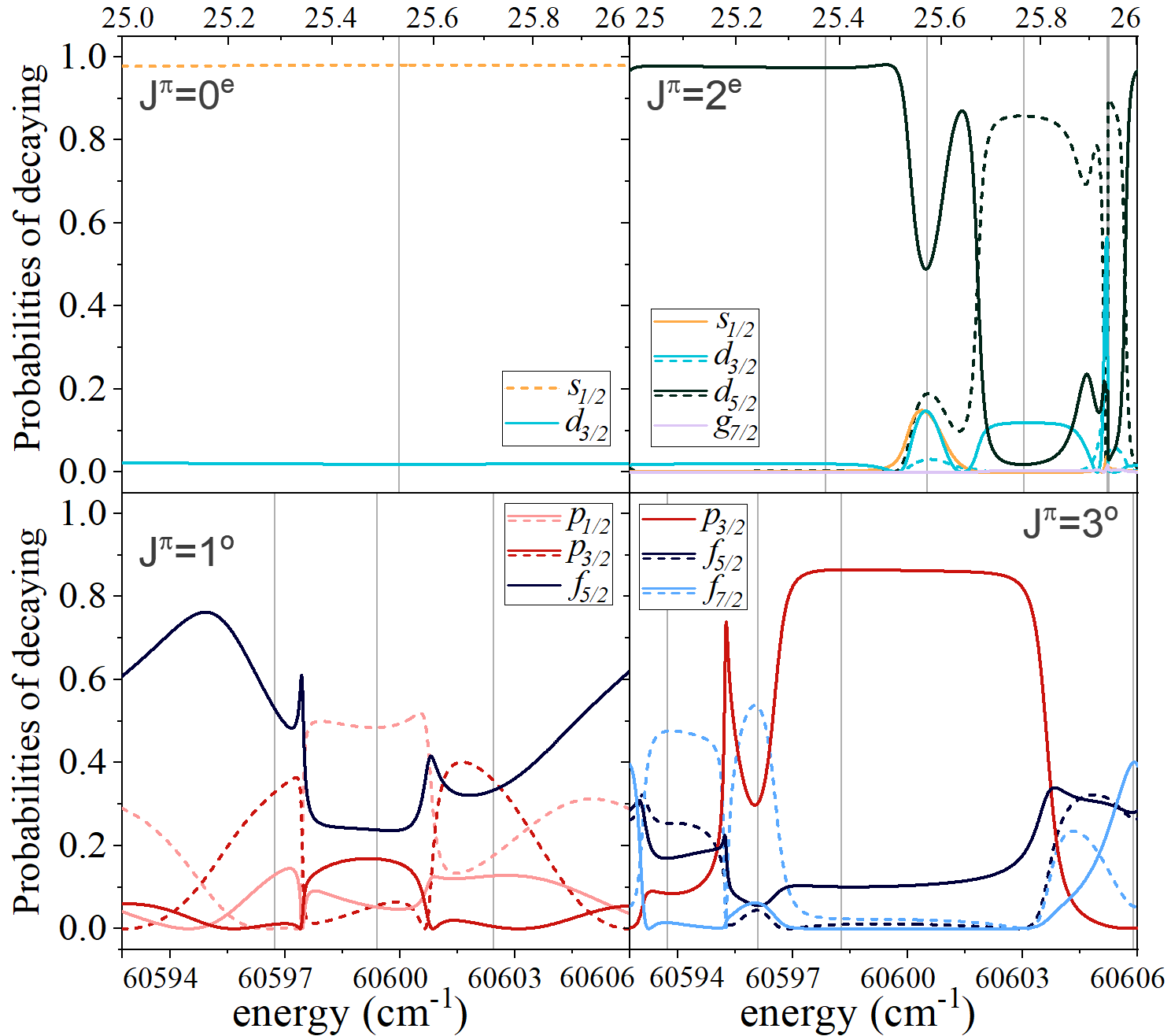}
  \caption{ Time delay analysis for $J_f^{\pi} = 0^e, 2^e$ (upper) and $1^o, 3^o$ (lower) resonances. Channels with different photoelectron angular momenta $l_j$ are distinguished by the colors, and components for the $5s_{1/2}$($4d_{3/2}$) threshold are given by dashed (solid) curves. The figures show the partial decay probabilities into each channel, obtained by computing the eigenvector associated with the dominant eigenvalue of the short-range time-delay matrix $\underline{Q}$. The vertical lines give the positions of the involved resonances.  }
  \label{5}
\end{figure}

As an attempt to figure out why the same resonance could have very different influences to $\delta_0^{(5s_{1/2})}$ and $\delta_0^{(4d_{3/2})}$, the time-delay matrices will be analyzed below, to determine the dominant decay channel in the autoionization process \cite{PhysRevA.45.1587,PhysRevA.103.033103}. The lifetime of a doubly-excited state can be estimated by the energy derivative of the physical scattering matrix $\underline{S}^{phys}$ \cite{Review:1996}:

\begin{equation}
\label{eq-5}
\underline{S}^{\dagger phys}=e^{-i\underline{\eta}}\underline{\tilde{S}}^{\dagger}e^{-i\underline{\eta}} \qquad
\underline{\tilde{S}}^{\dagger}= \underline{S}^{\dagger}_{oo}-\underline{S}^{\dagger}_{oc}(\underline{S}^{\dagger}_{cc}-e^{2i\underline{\beta}})^{-1}\underline{S}^{\dagger}_{co} 
\end{equation}
$\eta=\frac{\ln (2 k)}{k} + arg[\Gamma(l + 1 - \frac{i}{k})] - \frac{l\pi}{2}$ includes the Coulomb phase shift, which arises from the asymptotic Coulomb potential $-1/r$ and does not reveal the essence of electron correlations.  
When the final state energy $E=E_{thresh}+\frac{k^2}{2}$ is much higher above the ionization thresholds ($k\gtrsim 0.2$), 
the short-range unitary scattering matrix $\underline{\tilde{S}}$, obtained after closed channel elimination but without the Coulomb phaseshifts, can sometimes be used as an alternative to $\underline{S}^{phys}$ \cite{PhysRevA.45.1587}, since $d\eta/dE$ becomes nearly constant when $k\gg1$, and the energy derivatives of the two matrices give similar results. However, for the ionization of strontium, $Sr^{+}$ $4d_{3/2}$ and $4d_{5/2}$ thresholds are so close to each other that the energy we consider starts from only around 100 $cm^{-1}$ above the $4d_{3/2}$ threshold, where ${d\eta^{(4d_{3/2})}}/{dE}$ introduces strong oscillatory behaviors into $\underline{Q}^{phys}$:

\begin{equation}
\label{eq-6}
\begin{split}
    &\underline{Q}^{phys}=-i\underline{S}^{\dagger phys}\frac{d\underline{S}^{phys}}{dE}
    \qquad
    \underline{Q}=-i\underline{\tilde{S}}^{\dagger}\frac{d\underline{\tilde{S}}}{dE}\\
    &\underline{Q}^{phys}=\underline{S}^{\dagger phys}\frac{d\underline{\eta}}{dE}\underline{S}^{phys}
    +e^{-i\underline{\eta}}(\underline{Q}+\frac{d\underline{\eta}}{dE})e^{i\underline{\eta}}
\end{split}
\end{equation}

To get rid of the strong oscillatory features caused by $\eta$ (which we are not interested in), the short-range time-delay matrix $\underline{Q}$ is used to replace $\underline{Q}^{phys}$ in Fig. \ref{5}. The dominant decay channel analysis
 based on the two matrices is in fact the same for each resonance. 
Except for the $J^{\pi}=1^o$ symmetry whose spectrum does not display periodic behavior in the energy range of the figure, the dominant channels of decaying are different under this replacement.   
However, Fig. \ref{5}  cannot justify the relative strength of resonances to the two continua [the channels belong to $5s_{1/2}$ ($4d_{3/2}$) continuum are in dashed (solid) lines]. For example, in Fig. \ref{2}, $\delta_0^{(4d_{3/2})}$ varies with $J^{\pi}=0^e$ resonance in a much more dramatical way than $\delta_0^{(5s_{1/2})}$ , but according to Fig. \ref{5}, the $J^{\pi}=0^e$ resonance state should prominently decay into $5s_{1/2}$ rather than $4d_{3/2}$ continuum. In addition, both $J^{\pi}=3^o$ resonances near $\nu=25.1$ and $26$ have a larger influence to $\delta_0^{(4d_{3/2})}$ than $\delta_0^{(5s_{1/2})}$, but according to Fig. \ref{5} the two doubly-excited states actually decay prominently into $5s_{1/2}$ and $4d_{3/2}$ continuum respectively. Therefore the behaviors of these interference phases in each continuum remain difficult to interpret. 




\section{Conclusion}\label{s-4}

To conclude, we have studied the concurrent one- and two-photon ionization of an excited strontium atom, and have demonstrated its photoabsorption spectrum between the final state $4d_{3/2}$ and $4d_{5/2}$ thresholds. The positions of doubly excited states which have been reported by previous experiments are reproduced, and some resonances that have not yet been experimentally resolved are also predicted and identified. Moreover, the cross sections and angular distribution of photoelectrons are presented for each continuum and symmetry. The phase lag associated with the $\omega-2\omega$ coherent phase control at energies between the $4d_{3/2}$ and $4d_{5/2}$ thresholds are calculated following the method which successfully reproduced the phase lag observed in atomic barium \cite{PhysRevA.105.013113}. However, the phase lag we obtain for atomic strontium appears to be a less robust observable of this system, which is contrary to our expectation according to Ref. \cite{PhysRevA.105.013113}. A time-delay analysis is presented, but it could not enable a full interpretation of the phase lag behavior near each resonance. The influence of long-range Coulomb potential on the lifetimes of the double excited states has been discussed in the energy range close to the $4d_{3/2}$ threshold.

\begin{acknowledgments}
This work was supported by the U.S. Department of Energy, Office of Science, Basic Energy Sciences, under Award No. DE-SC0010545.
\end{acknowledgments}

\appendix*
\section{\label{sec:qdt} Quantum-defect matrices}

In Table. \ref{tab:table1} we give the fragmentation channels involved and the calculated quantum defect matrices (whose definition refers to Ref. \cite{PhysRevA.36.2706,PhysRevA.36.4272}) for all the symmetries $J^{\pi}$. The energy is at $\nu=25$ relative to the $4d_{5/2}$ threshold. With this table, it should be possible to reproduce most of our results in the energy range considered.

\begin{table*}
\caption{\label{tab:table1}}
\begin{ruledtabular}
\begin{tabular}{cccccccccccc}
\multicolumn{12}{c}{Quantum-defect matrices $\mu_{ij}$ at $\nu=25$ with $R$-matrix box 70 $a.u.$.  }\\
 \hline
$0^e$ & $5s_{1/2}\epsilon s_{1/2}$ & $4d_{3/2}\epsilon d_{3/2}$ & $4d_{5/2}\epsilon d_{5/2}$ & & & & & & & \\ 
 \hline
 & 0.2576 & -0.0260 & -0.0433 & & & & & & & \\
 & -- & 0.4626 & 0.0077 & & & & & & & \\
 & -- & -- & 0.4678 & & & & & & & \\
 \hline
$1^e$ & $5s_{1/2}\epsilon s_{1/2}$ & $5s_{1/2}\epsilon d_{3/2}$ &  $4d_{3/2}\epsilon s_{1/2}$ & $4d_{3/2}\epsilon d_{3/2}$ & $4d_{3/2}\epsilon d_{5/2}$ & $4d_{5/2}\epsilon d_{3/2}$ & $4d_{5/2}\epsilon d_{5/2}$ & $4d_{5/2}\epsilon g_{7/2}$ & & & \\
 \hline
 & 0.3289 & -0.2d-4 & 0.4d-4 & -0.0050 & -0.0106 & 0.0106 & 0.0103 & 0.3d-4 & & & \\
 & -- & -0.2540 & 0.1030 & -0.0123 & 0.0051 & -0.0077 & 0.0068 & -0.0074 & & & \\
 & -- & -- & 0.2423 & -0.0044 & 0.0049 & -0.0019 & 0.0046 & 0.0104 & & & \\
 & -- & -- & -- & -0.2007 & -0.0099 & 0.0008 & -0.0008 & -0.0007 & & & \\
 & -- & -- & -- & -- & 0.0891 & 0.3319 & 0.0115 & 0.0019 & & & \\
 & -- & -- & -- & -- & -- & 0.1331 & -0.0039 & 0.0003 & & & \\
 & -- & -- & -- & -- & -- & -- & -0.2248 & 0.0013 & & & \\
 & -- & -- & -- & -- & -- & -- & -- & 0.0607 & & & \\
 \hline
$2^e$ & $5s_{1/2}\epsilon d_{3/2}$ & $5s_{1/2}\epsilon d_{5/2}$ & $4d_{3/2}\epsilon s_{1/2}$ & $4d_{3/2}\epsilon d_{3/2}$ & $4d_{3/2}\epsilon d_{5/2}$ & $4d_{3/2}\epsilon g_{7/2}$ & $4d_{5/2}\epsilon s_{1/2}$ & $4d_{5/2}\epsilon d_{3/2}$ & $4d_{5/2}\epsilon d_{5/2}$ & $4d_{5/2}\epsilon g_{7/2}$ & $4d_{5/2}\epsilon g_{9/2}$ \\
 \hline
 & -0.2663 & 0.0152 & -0.1170 & 0.0180 & 0.0018 & 0.0119 & -0.0116 & -0.0172 & 0.0427 & -0.0046 & 0.0066 \\
 & -- & -0.2777 & 0.0187 & -0.0175 & -0.0162 & -0.0073 & 0.1177 & -0.0030 & -0.0491 & -0.0010 & -0.0154 \\
 & -- & -- & 0.2280 & -0.0093 & -0.0030 & 0.0125 & -0.0167 & 0.0132 & -0.0167 & -0.0063 & 0.0023 \\
 & -- & -- & -- & -0.1970 & 0.2520 & -0.0161 & -0.0155 & -0.2635 & 0.1742 & -0.0218 & 0.0087 \\
 & -- & -- & -- & -- & 0.0093 & 0.0148 & -0.0210 & -0.2224 & -0.0729 & 0.0041 & -0.0023 \\
 & -- & -- & -- & -- & -- & 0.0942 & 0.0061 & -0.0126 & 0.0132 & 0.0049 & -0.0031 \\
 & -- & -- & -- & -- & -- & -- & 0.2134 & 0.0117 & -0.0267 & 0.0016 & 0.0119 \\
 & -- & -- & -- & -- & -- & -- & -- & 0.0306 & 0.0636 & -0.0061 & -0.0001 \\
 & -- & -- & -- & -- & -- & -- & -- & -- & 0.3599 & 0.0027 & 0.0014 \\
 & -- & -- & -- & -- & -- & -- & -- & -- & -- & 0.0621 & -0.0004 \\
 & -- & -- & -- & -- & -- & -- & -- & -- & -- & -- & 0.0590 \\
 \hline
$1^o$ & $5s_{1/2}\epsilon p_{1/2}$ & $5s_{1/2}\epsilon p_{3/2}$ & $4d_{3/2}\epsilon p_{1/2}$ & $4d_{3/2}\epsilon p_{3/2}$ & $4d_{3/2}\epsilon f_{5/2}$ & $4d_{5/2}\epsilon p_{3/2}$ & $4d_{5/2}\epsilon f_{5/2}$ & $4d_{5/2}\epsilon f_{7/2}$  &  &  & \\ 
 \hline
 & -0.0313 & 0.0497 & -0.0165 & -0.0658 & -0.1177 & -0.0568 & 0.1784 & 0.0191 &  &  & \\ 
 & -- & -0.1102 & 0.0217 & 0.0044 & 0.0290 & 0.0251 & 0.0536 & 0.1454 &  &  & \\ 
 & -- & -- & -0.1731 & 0.0663 & 0.0903 & 0.0220 & -0.0691 & 0.0516 &  &  & \\ 
 & -- & -- & -- & -0.1168 & 0.0835 & -0.0126 & -0.1490 & -0.0722 &  &  & \\ 
 & -- & -- & -- & -- & 0.0033 & 0.1185 & -0.0676 & -0.0634 &  &  & \\ 
 & -- & -- & -- & -- & -- & -0.0532 & -0.0374 & 0.2264 &  &  & \\ 
 & -- & -- & -- & -- & -- & -- & 0.1614 & 0.1181 &  &  & \\ 
 & -- & -- & -- & -- & -- & -- & -- & 0.0502 &  &  & \\ 
 \hline
$2^o$ & $5s_{1/2}\epsilon p_{3/2}$ & $5s_{1/2}\epsilon f_{5/2}$ & $4d_{3/2}\epsilon p_{1/2}$ & $4d_{3/2}\epsilon p_{3/2}$ & $4d_{3/2}\epsilon f_{5/2}$ & $4d_{3/2}\epsilon f_{7/2}$ & $4d_{5/2}\epsilon p_{1/2}$ & $4d_{5/2}\epsilon p_{3/2}$ & $4d_{5/2}\epsilon f_{5/2}$ & $4d_{5/2}\epsilon f_{7/2}$ & \\
 \hline
 & -0.2081 & -0.0002 & -0.0097 & -0.0245 & 0.0246 & 0.0787 & 0.0408 & 0.0500 & -0.0472 & -0.0847 & \\
 & -- & 0.1281 & -0.0528 & 0.0234 & -0.0125 & 0.0040 & -0.0120 & 0.0119 & -0.0154 & 0.0074 & \\
 & -- & -- & -0.1749 & -0.0095 & -0.0299 & 0.0075 & -0.0031 &  0.0132 & -0.0209 & -0.0048 & \\
 & -- & -- & -- & -0.2262 & 0.0074 & 0.0137 & -0.0104 & 0.0332 & -0.0114 & -0.0332 & \\
 & -- & -- & -- & -- & 0.0365 & -0.0534 & 0.0016 & -0.0148 & 0.0550 & 0.0433 & \\
 & -- & -- & -- & -- & -- & -0.1394 & 0.0099 & -0.0305 & 0.0718 & 0.1629 & \\
 & -- & -- & -- & -- & -- & -- & -0.2127 & -0.0217 & -0.0022 & 0.0271 & \\
 & -- & -- & -- & -- & -- & -- & -- & -0.2301 & 0.0322 & 0.0226 & \\
 & -- & -- & -- & -- & -- & -- & -- & -- & -0.0660 & -0.1020 & \\
 & -- & -- & -- & -- & -- & -- & -- & -- & -- & -0.1626 & \\
 \hline
$3^o$ & $5s_{1/2}\epsilon f_{5/2}$ & $5s_{1/2}\epsilon f_{7/2}$ & $4d_{3/2}\epsilon p_{3/2}$ & $4d_{3/2}\epsilon f_{5/2}$ & $4d_{3/2}\epsilon f_{7/2}$ & $4d_{5/2}\epsilon p_{1/2}$ & $4d_{5/2}\epsilon p_{3/2}$ & $4d_{5/2}\epsilon f_{5/2}$ & $4d_{5/2}\epsilon f_{7/2}$ &  & \\
 \hline
 & 0.0984 & 0.0292 & -0.0701 & 0.0286 & 0.0011 & 0.0023 & -0.0194 & -0.0384 & 0.0315 &  & \\
 & -- & 0.0945 & 0.0293 & -0.0163 & -0.0181 & -0.0805 & 0.0457 & 0.0050 & -0.0328 &  & \\
 & -- & -- & -0.2691 & -0.0362 & -0.0167 & 0.0734 & -0.0612 & 0.0296 & -0.0633 &  & \\
 & -- & -- & -- & 0.0184 & -0.0174 & 0.0382 & -0.0422 & 0.0504 & -0.0485 &  & \\
 & -- & -- & -- & -- & 0.0375 & -0.0172 & -0.0019 & 0.0272 & 0.0188 &  & \\
 & -- & -- & -- & -- & -- & -0.2497 & 0.0397 & -0.0614 & 0.0414 &  & \\
 & -- & -- & -- & -- & -- & -- & -0.3103 & 0.0413 & -0.0776 &  & \\
 & -- & -- & -- & -- & -- & -- & -- & -0.0125 & 0.0340 &  & \\
 & -- & -- & -- & -- & -- & -- & -- & -- & -0.0911 &  & \\
\end{tabular}
\end{ruledtabular}
\end{table*}


\bibliographystyle{unsrt}
\bibliography{Sr}

\end{document}